\documentclass[a4paper,11pt]{article}

\usepackage{pos}
\usepackage{lipsum} 

\title{Search for Quantum Decoherence with 10.7 years of atmospheric $\nu_\mu$ events in IceCube}

\ShortTitle{Quantum Decoherence in IceCube}

\author{The IceCube Collaboration \\{\normalsize \normalfont(a complete list of authors can be found at the end of the proceedings)}\\}

\emailAdd{tkrishnan@g.harvard.edu}

\abstract{

In order to develop a consistent quantum theory of gravity, we must understand the nature of spacetime at the Planck scale. In particular if it exhibits quantum fluctuations, they may cause propagating particles to evolve in an apparently non-unitary manner. Neutrinos, which interact only via the weak force and gravity, maintain quantum coherence while propagating over large distances. Thus, neutrino oscillations serve as a precise interferometer to search for Planck-scale fluctuations of spacetime. The IceCube Neutrino Observatory is the world’s largest neutrino telescope, located in the Antarctic icecap. We search the data on atmospheric neutrinos detected by IceCube in the energy range 0.5-100 TeV to test for neutrino decoherence. In this contribution, we present the sensitivity of the analysis, which shows significant improvement compared to previous IceCube results as a result of improved reconstruction and a larger sample of events.

\vspace{4mm}

{\bfseries Corresponding authors:}

Tanvi Krishnan$^{1*}$\\

{$^{1}$ \itshape Harvard University}\\[4mm]
$^*$ Presenter
}

\ConferenceLogo{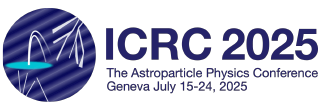}

\FullConference{39th International Cosmic Ray Conference (ICRC2025)\\
 15–24 July 2025\\
Geneva, Switzerland\\}

\begin{document}

\maketitle

\section{Introduction}

The possible quantum nature of spacetime and gravity at very high energies approaching the Planck scale is difficult to probe experimentally. Since neutrinos interact only very weakly with matter and maintain their quantum coherence over large distances, neutrino oscillations serve as a precise interferometer to search for Planck-scale fluctuations of spacetime, which would cause apparent violation of unitarity \cite{stuttard_qg}. In order to study this, we need a well-understood beam of neutrinos propagating over long baselines.

We use an atmospheric muon neutrino sample from the IceCube Neutrino Observatory to search for this signature. A previous search with 7.6 years of data has produced the strongest experimental constraints on neutrino decoherence due to quantum gravity to date \cite{gp_deco_paper}. We present the latest developments on an updated search with 10.7 years of data and improved treatment of systematics. 

\section{The IceCube Detector}
The IceCube Neutrino Observatory \cite{ic_detector} is a detector that encloses a cubic kilometer of glacial ice at the South Pole. It consists of 5,160 Digital Optical Modules (DOMs), each containing a photomultiplier (PMT), distributed among 86 strings. The strings are spaced 125 meters apart. The DOMs record Cherenkov light emitted along the path of relativistic charged particles produced in neutrino interactions. Reconstruction allows us to calculate properties of the primary particles (energy and direction). There is also a more densely-instrumented region of 8 strings, called DeepCore, at the center of the detector, which is useful for reconstruction of low-energy events. 

\section{Quantum Decoherence}

\begin{figure}[ht]
    \centering
    \includegraphics[width=0.5\linewidth]{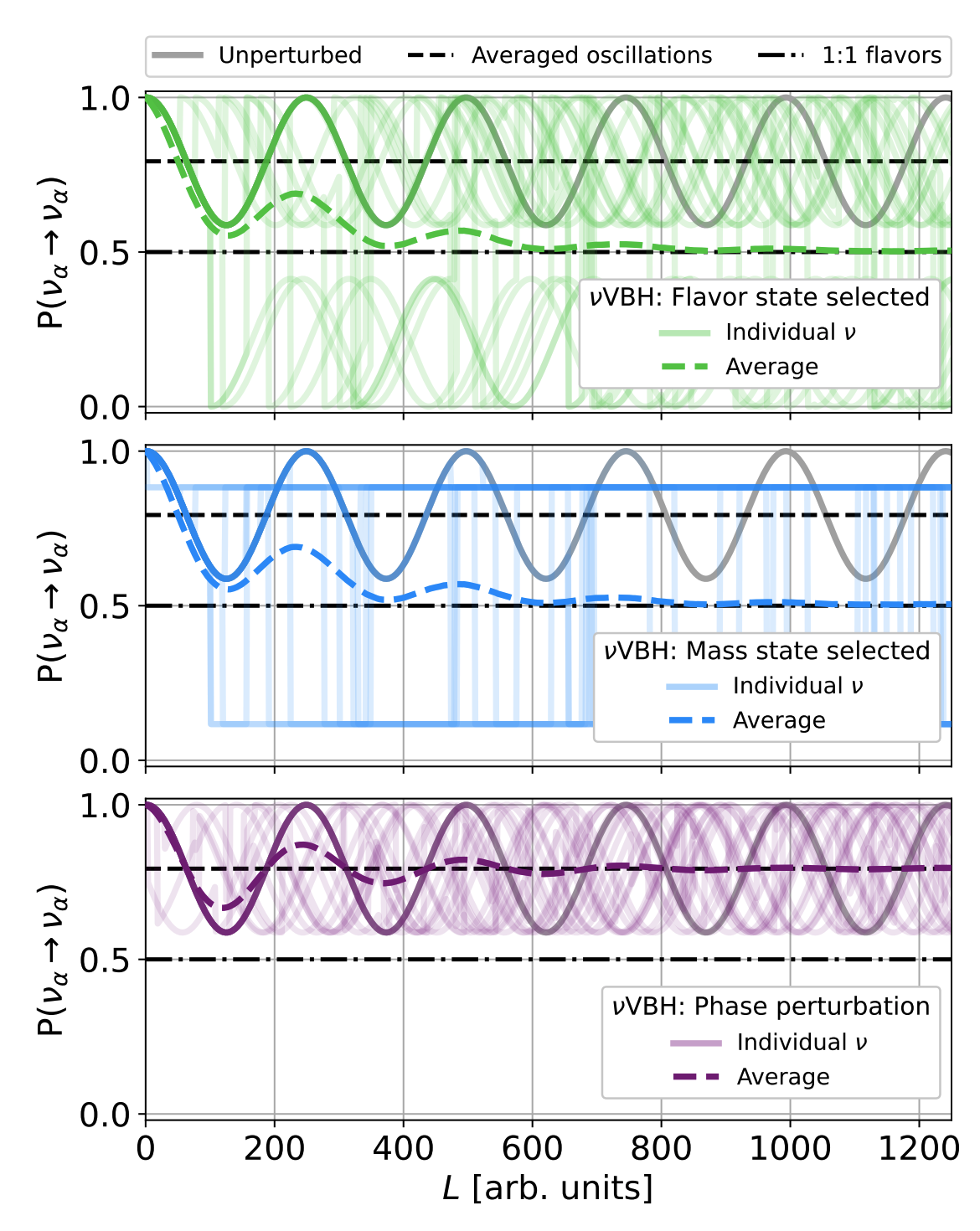}
    \caption{Decoherence of an oscillating neutrino ensemble. The emergence of non-unitary behavior from various types of interactions between neutrinos and VBHs. The top two rows correspond to the state selection model where the neutrino wavefunction is collapsed into a democratically selected flavor or mass eigenstate. The bottom row corresponds to the phase perturbation model, where the neutrino wavefunction's phase in the mass basis is perturbed. Taken from \cite{gp_deco_paper}}
    \label{fig:model}
\end{figure}
Oscillations of atmospheric neutrinos are well understood experimentally \cite{superK}, since neutrinos interact only very weakly with matter. Thus, neutrino wavefunctions can propagate in a coherent manner over large distances. However, if there exist even very weak Planck-scale fluctuations of spacetime, over large distances these may destroy the coherence of the oscillating neutrinos. We focus on neutrino interactions with virtual black holes (VBHs) produced by spacetime fluctuations \cite{hawking}. We study two different heuristic models for these interactions: democratic mass/flavor state selection and mass state phase perturbation. We will refer to these two interaction models as \textit{state selection} and \textit{phase perturbation}, respectively. The effect of these two models on an ensemble of propagating neutrinos can be seen in \autoref{fig:model}.

\begin{figure}[ht]
    \centering
    \includegraphics[width=0.5\linewidth]{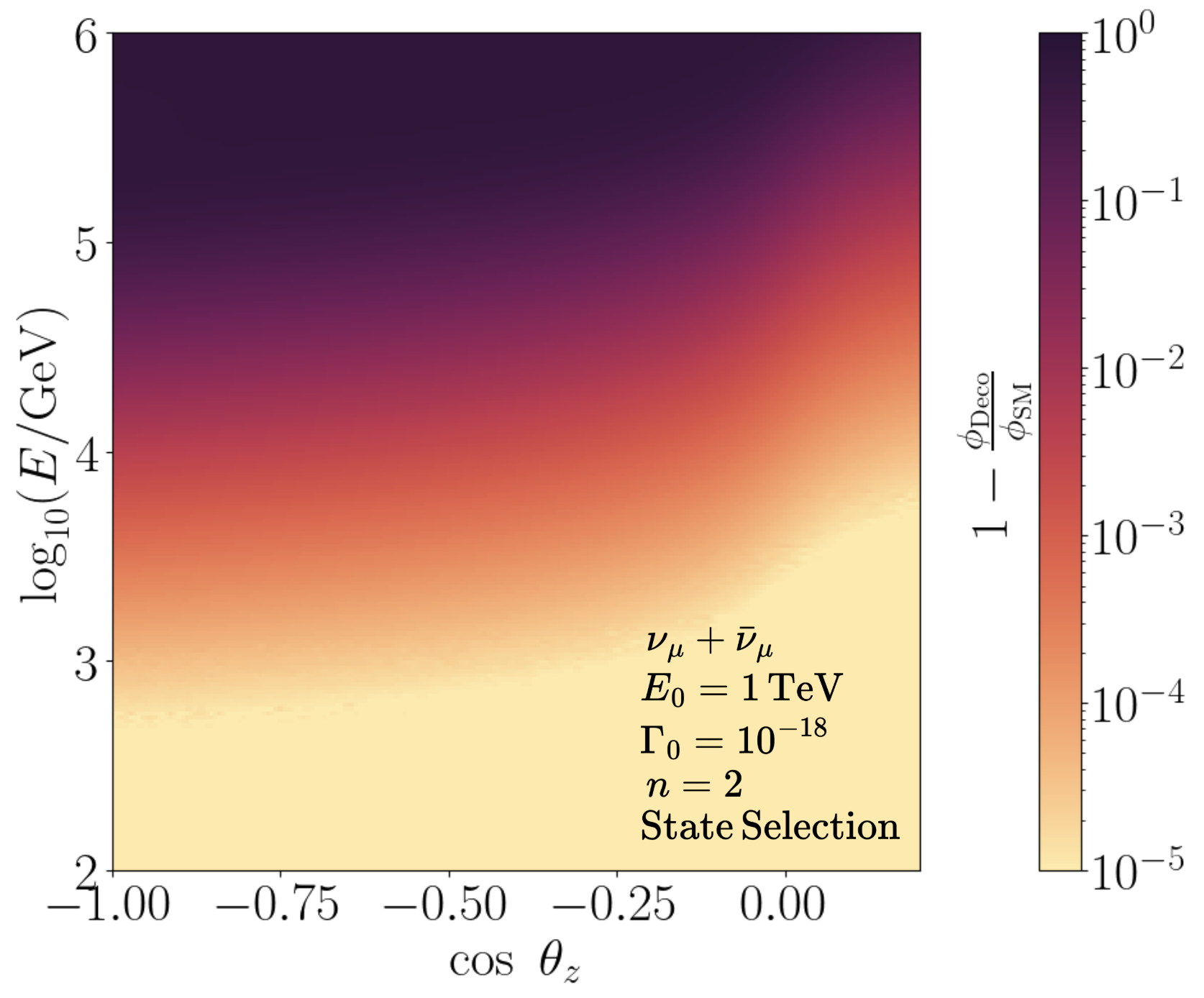}
    \caption{Muon neutrino oscillogram.  Ratio of the predicted $\nu_\mu$ flux for the state selection model with the decoherence parameters listed compared to the standard model prediction.}
    \label{fig:oscillogram}
\end{figure}

We use the formalism of open quantum systems to encode the time evolution of a neutrino coupled to its environment, using the Lindblad master equation (in natural units with $\hbar=c=1$):
\begin{equation}
\dot{\rho}=-i[H,\rho]-\mathcal{D}[\rho]
\end{equation}
where $H$ is the Hamiltonian of the neutrino system and $\mathcal{D}[\rho]$ is the decoherence superoperator \cite{openquantumsys}. The decoherence superoperator is dependent on the underlying physics causing the decoherence effect. For a three-neutrino system, it can be expanded in the basis of the Gell-Mann matrices ($b^\mu$): 
\begin{equation}
\mathcal{D}[\rho]=(D_{\mu\nu}\rho^{\nu})b^{\mu}
\end{equation} 
where $\rho^{\nu}$ are the coefficients of the system's density matrix expanded in the basis of $b^{\mu}$. The matrix $D_{\mu\nu}$ is an $9\times 9$ matrix that parametrizes the decoherence effects in a chosen model. The state selection model, in the long-distance limit, predicts a flux that is equally weighted in all neutrino flavors, independent of the basis of randomization. The phase perturbation model, in the long-distance limit, predicts a flux that tends towards an incoherent sum of mass eigenstates. The simplest way to encode these averaging behaviors is by taking $D$ as diagonal matrices parametrized by a single parameter $\Gamma$:
\begin{equation}
    D_{\text{state selection}} = \text{diag}(0, \Gamma, \Gamma, \Gamma, \Gamma, \Gamma, \Gamma,\Gamma, \Gamma)
\end{equation}
\begin{equation}
D_{\text{phase perturbation}} = \text{diag}(0, \Gamma, \Gamma, 0, \Gamma, \Gamma, \Gamma,\Gamma, 0)
\end{equation}
where we assume that $\Gamma$ has a power law energy dependence:
\begin{equation}
    \Gamma(E) = \Gamma_0\big(\frac{E}{E_0}\big)^n
\end{equation}
where $E_0$ is a reference energy. Thus, our model has two free parameters: $n$ and $\Gamma_0$. The parameter $n$ denotes the energy-dependence of decoherence, while $\Gamma_0$ denotes the strength of decoherence at the reference energy. An example oscillogram for a representative set of parameters is shown in\autoref{fig:oscillogram}, showing the change in $\nu_\mu$ flux as a function of energy and zenith angle across the high-energy IceCube $\nu_\mu$ sample.

\section{Event Selection and Systematic Uncertainties}
This analysis constrains $\nu\text{VBH}$ models with 10.7 years of IceCube data using $368,071 $ up-going $\nu_\mu$ events with reconstructed energies ranging from $0.5$ to $100$ TeV. The updated event selection is described in detail in \cite{alfonso_sterile} and has been used by the IceCube collaboration for searches for eV-scale sterile neutrinos, and other searches currently in progress. 

The $\nu\text{VBH}$ interaction signature was previously searched for with 7.6 years of IceCube data \cite{gp_deco_paper}, but this new analysis has not only a larger livetime but also a higher purity (increased from 99\% to 99.9\%). The updated analysis also has improved treatment of systematics. These include parameters associated with the conventional and non-conventional neutrino fluxes, normalization, ice properties, detector response, and neutrino attenuation. Each uncertainty contribution is treated as a continuous nuisance parameter in a likelihood maximization analysis, comparing the best-fit likelihood at many values of $\Gamma$ to establish a frequentist confidence interval for each energy dependence exponent $n$ and decoherence model. This new analysis also separates starting vs. throughgoing tracks and has an improved BDT selection for final-level cuts. 

\section{Analysis Method and Sensitivity}
\begin{figure}[ht]
    \centering
    \includegraphics[width=0.7\linewidth]{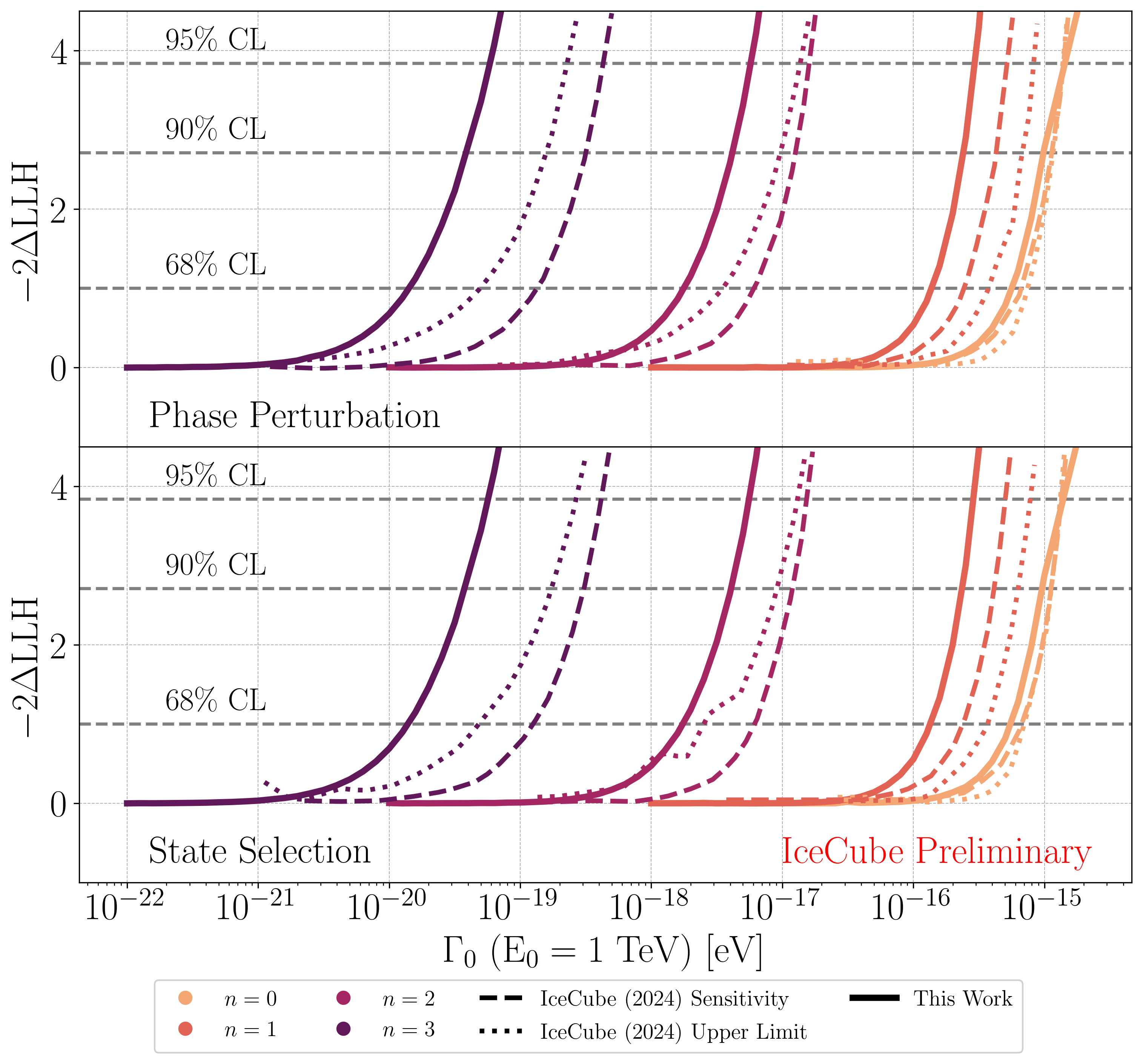}
    \caption{Sensitivities of the old (dashed) \cite{gp_deco_paper} and new (solid) analyses for both models: phase perturbation (top) and state selection (bottom). The prior analysis's upper limit is also shown (dotted). The sensitivity has greatly improved since the last iteration of the analysis.}
    \label{fig:sensitivities}
\end{figure}

The likelihood test statistic accounts for both data and Monte Carlo statistical uncertainties, following Ref.  \cite{carlos_frequentist}. We constrain $\Gamma_0$ for $n=0, 1, 2, 3$ using events separated into starting and through-going (events with vertices inside and outside the instrumented detector volume, respectively) which are each binned logarithmically into 24 bins in reconstructed muon energy, $\log(E^\mu_{\text{reco}})$, where $E^\mu_{\text{reco}}\in[500\,\text{GeV}, 100\,\text{TeV}]$ and uniformly in 18 bins in zenith angle, $\cos(\theta^\mu_{\text{reco}})$, where $\cos(\theta^\mu_{\text{reco}})\in[-1.0, -0.1]$. The Asimov sensitivities are shown in \autoref{fig:sensitivities} in comparison to the sensitivities and upper limits obtained from the previous analysis \cite{gp_deco_paper}. From this, we see that our sensitivities are greatly improved compared to the prior analysis.

\section{Conclusions}
An updated quantum decoherence analysis using improved systematic treatment and 10.7 years of upward-going muon neutrino events has been presented. This analysis has world-leading sensitivity to quantum  decoherence. Having verified the expected sensitivity using simulations, we are now proceeding towards a measurement using unblinded data.

\bibliographystyle{ICRC}
\bibliography{ICRC2025_template_IceCube}

\clearpage

\section*{Full Author List: IceCube Collaboration}

\scriptsize
\noindent
R. Abbasi$^{16}$,
M. Ackermann$^{63}$,
J. Adams$^{17}$,
S. K. Agarwalla$^{39,\: {\rm a}}$,
J. A. Aguilar$^{10}$,
M. Ahlers$^{21}$,
J.M. Alameddine$^{22}$,
S. Ali$^{35}$,
N. M. Amin$^{43}$,
K. Andeen$^{41}$
% ...
\\
\\
% ...
$^{10}$ Universit{\'e} Libre de Bruxelles, Science Faculty CP230, B-1050 Brussels, Belgium \\
% ...
$^{16}$ Department of Physics, Loyola University Chicago, Chicago, IL 60660, USA \\
$^{17}$ Dept. of Physics and Astronomy, University of Canterbury, Private Bag 4800, Christchurch, New Zealand \\
% ...
$^{21}$ Niels Bohr Institute, University of Copenhagen, DK-2100 Copenhagen, Denmark \\
$^{22}$ Dept. of Physics, TU Dortmund University, D-44221 Dortmund, Germany \\
% ...
$^{35}$ Dept. of Physics and Astronomy, University of Kansas, Lawrence, KS 66045, USA \\
% ...
$^{39}$ Dept. of Physics and Wisconsin IceCube Particle Astrophysics Center, University of Wisconsin{\textemdash}Madison, Madison, WI 53706, USA \\
% ...
$^{41}$ Department of Physics, Marquette University, Milwaukee, WI 53201, USA \\
% ...
$^{43}$ Bartol Research Institute and Dept. of Physics and Astronomy, University of Delaware, Newark, DE 19716, USA \\
% ...
$^{63}$ Deutsches Elektronen-Synchrotron DESY, Platanenallee 6, D-15738 Zeuthen, Germany \\
$^{\rm a}$ also at Institute of Physics, Sachivalaya Marg, Sainik School Post, Bhubaneswar 751005, India \\
% ...

\subsection*{Acknowledgments}

\noindent
This material is based upon work supported by the U.S. Department of Energy, Office of Science, Office of Advanced Scientific Computing Research, Department of Energy Computational Science Graduate Fellowship. The authors gratefully acknowledge the support from the following agencies and institutions:
USA {\textendash} U.S. National Science Foundation-Office of Polar Programs,
U.S. National Science Foundation-Physics Division,
U.S. National Science Foundation-EPSCoR,
U.S. National Science Foundation-Office of Advanced Cyberinfrastructure,
Wisconsin Alumni Research Foundation,
Center for High Throughput Computing (CHTC) at the University of Wisconsin{\textendash}Madison,
Open Science Grid (OSG),
Partnership to Advance Throughput Computing (PATh),
Advanced Cyberinfrastructure Coordination Ecosystem: Services {\&} Support (ACCESS),
Frontera and Ranch computing project at the Texas Advanced Computing Center,
U.S. Department of Energy-National Energy Research Scientific Computing Center,
Particle astrophysics research computing center at the University of Maryland,
Institute for Cyber-Enabled Research at Michigan State University,
Astroparticle physics computational facility at Marquette University,
NVIDIA Corporation,
and Google Cloud Platform;
Belgium {\textendash} Funds for Scientific Research (FRS-FNRS and FWO),
FWO Odysseus and Big Science programmes,
and Belgian Federal Science Policy Office (Belspo);
Germany {\textendash} Bundesministerium f{\"u}r Forschung, Technologie und Raumfahrt (BMFTR),
Deutsche Forschungsgemeinschaft (DFG),
Helmholtz Alliance for Astroparticle Physics (HAP),
Initiative and Networking Fund of the Helmholtz Association,
Deutsches Elektronen Synchrotron (DESY),
and High Performance Computing cluster of the RWTH Aachen;
Sweden {\textendash} Swedish Research Council,
Swedish Polar Research Secretariat,
Swedish National Infrastructure for Computing (SNIC),
and Knut and Alice Wallenberg Foundation;
European Union {\textendash} EGI Advanced Computing for research;
Australia {\textendash} Australian Research Council;
Canada {\textendash} Natural Sciences and Engineering Research Council of Canada,
Calcul Qu{\'e}bec, Compute Ontario, Canada Foundation for Innovation, WestGrid, and Digital Research Alliance of Canada;
Denmark {\textendash} Villum Fonden, Carlsberg Foundation, and European Commission;
New Zealand {\textendash} Marsden Fund;
Japan {\textendash} Japan Society for Promotion of Science (JSPS)
and Institute for Global Prominent Research (IGPR) of Chiba University;
Korea {\textendash} National Research Foundation of Korea (NRF);
Switzerland {\textendash} Swiss National Science Foundation (SNSF).

\end{document}